\documentclass[cits]{PoS}\usepackage{amsmath,amsbsy,cite}

\title{Tetraquark Properties at Large $N_{\rm c}$} 
\ShortTitle{Tetraquark Properties at Large $N_{\rm c}$}
\author{Wolfgang Lucha\\Institute for High Energy Physics, Austrian
Academy of Sciences, Nikolsdorfergasse 18,\\A-1050 Vienna,
Austria\\E-mail: \email{Wolfgang.Lucha@oeaw.ac.at}}
\author{Dmitri Melikhov\\Institute for High Energy
Physics, Austrian Academy of Sciences, Nikolsdorfergasse
18,\\A-1050 Vienna, Austria, and\\D.~V.~Skobeltsyn Institute of
Nuclear Physics, M.~V.~Lomonosov Moscow State University,\\119991,
Moscow, Russia, and\\Faculty of Physics, University of Vienna,
Boltzmanngasse 5, A-1090 Vienna, Austria\\E-mail:
\email{dmitri\_melikhov@gmx.de}}\author{\speaker{Hagop
Sazdjian}\\Institut de Physique Nucl\'eaire, Universit\'e
Paris-Sud, CNRS-IN2P3, Universit\'e Paris-Saclay,\\91405 Orsay
Cedex, France\\E-mail: \email{sazdjian@ipno.in2p3.fr}}
\abstract{Considering quantum chromodynamics in the consistent
limit of simultaneous \emph{correlated\/} increase of the number
of colours, $N_{\rm c}$, beyond bounds and decrease of the strong
coupling to zero allows for solid statements about qualitative
features of the class of ``truly exotic'' tetraquark mesons
carrying four mutually distinct quark flavours. Consistency
criteria extracted from correlation functions for
two-ordinary-meson scattering suggest the existence of more than
one such tetraquarks, at~least, of two tetraquarks of identical
quark-flavour content and large-$N_{\rm c}$ behaviour of the total
decay~widths but differing in, and hence discriminable by, their
predominant decay modes into two conventional mesons. This
pairwise appearance is in conflict with the unique variant of such
a four-quark bound state arising from the binding of
\emph{diquark\/} and \emph{antidiquark\/} to a tetraquark by the
strong interactions.}

\FullConference{Light Cone 2019 --- QCD on the light cone: from
hadrons to heavy ions --- LC2019\\16--20 September 2019\\\'Ecole
Polytechnique, Palaiseau, France}

\begin{document}\section{Approach to Non-Conventional Hadrons:
Multiquark Hadrons in Large-$N_{\rm c}$ QCD}Quantum
chromodynamics, the quantum field theory of the strong
interactions, does not forbid the formation of hadrons
(colour-singlet bound states of quarks and gluons) other than
conventional quark--antiquark mesons $(\overline q\,q)$ and
three-quark baryons $(q\,q\,q)$, \emph{viz.}, of exotic
\emph{multiquark\/} states such as tetraquark $(\overline
q\,\overline q\,q\,q)$, pentaquark $(\overline q\,q\,q\,q\,q)$,
and hexaquark $(q\,q\,q\,q\,q\,q)$ or $(\overline q\,\overline
q\,\overline q\,q\,q\,q)$ hadron states. The prospects for such
exotics have been studied since long \cite{RJ}. We focus to the
case of tetraquarks.

The framework of our analysis is a generalization of quantum
chromodynamics (QCD) dubbed large-$N_{\rm c}$ QCD \cite{GH}: a
gauge theory relying on the gauge group ${\rm SU}(N_{\rm c})$,
with all quarks transforming, \emph{by assumption\/}, according to
the $N_{\rm c}$-dimensional, fundamental representation of ${\rm
SU}(N_{\rm c})$, considered in the limit $N_{\rm c}\to\infty$,
with the strong coupling parameter, $g_{\rm s}$, behaving like
$g_{\rm s}\propto1/{N_{\rm c}^{1/2}}$. In that limit, QCD catches
the main properties of confinement, while getting simplified with
respect to secondary complications, \emph{e.g.}, inelasticity or
screening effects. $1/{N_{\rm c}}$ plays the r\^ole of a
perturbative~parameter. The theory's predictions are easily
deduced \cite{EW}: the spectrum is saturated by an infinite tower
of~free stable mesons with masses of order $O(N_{\rm c}^0)$,
three-meson interactions behave like $N_{\rm c}^{-1/2}$,
four-meson interactions like $N_{\rm c}^{-1}$, and meson decay
rates like $N_{\rm c}^{-1}$. So, in the limit $N_{\rm c}\to\infty$
all mesons are stable, which is one of the main features of
confinement. Can we derive similar predictions for~tetraquarks?

At large $N_{\rm c}$, adopting the two-point correlators of
colour-singlet tetraquark and ordinary-meson currents, the
propagation of a tetraquark becomes equivalent to that of two free
ordinary mesons \cite{C}:$$T(x)\equiv(\overline q\,\overline q\,q
\,q)(x)\ ,\quad j(x)\equiv(\overline q\,q)(x)\qquad\Longrightarrow
\qquad\langle T(x)\,T^\dag(0)\rangle\stackrel{N_{\rm c}\to\infty}
{=}\sum\langle j(x)\,j^\dag(0)\rangle\,\langle j(x)\,j^\dag(0)
\rangle\ .$$Tetraquark poles thus cannot appear at leading order
but may pop up at subleading orders and~might be observable if
their widths turn out to be narrow; the latter are expected to
fall off like $N_{\rm c}^{-2}$ \cite{SW,KP,CL,MPR1,TQN1,TQN2}.

\section{Tetraquarks of Flavour-Exotic Quark--Antiquark
Composition: \emph{Line of Approach}} We are particularly
interested in flavour-exotic tetraquarks: bound states of two
quarks and two antiquarks involving four different quark flavours,
here generically denoted $1,2,3,4\in\{u,d,s,c,b\}$.

We extract basic features of tetraquarks from their appearance as
poles in the amplitudes for the scattering of two ordinary mesons
into two ordinary mesons by an investigation \cite{TQN1,TQN2,TQN3}
of four-point Green functions of colour-singlet quark-bilinear
currents, serving as meson interpolating operators. Identifying
spin and parity as irrelevant for \emph{qualitative\/} analyses,
such a current $j_{ab}$ generically~reads$$j_{ab}\equiv\overline
q_a\,q_b\ ,\qquad a,b=1,2,3,4\ ,$$ and couples to an ordinary
meson $M_{ab}=(\overline q_a\,q_b)$ with strength $f_{M_{ab}}$ of
known large-$N_{\rm c}$ behaviour~\cite{EW}:$$\langle0|j_{ab}
|M_{ab}\rangle=f_{M_{ab}}\ ,\qquad f_{M_{ab}}\propto N_{\rm
c}^{1/2}\ .$$Consistency requires us to inspect all channels with
potential tetraquark poles, by use of the currents$$j_{12}\equiv
\overline q_1\,q_2\ ,\qquad j_{34}\equiv\overline q_3\,q_4\
,\qquad j_{14}\equiv\overline q_1\,q_4\ ,\qquad j_{32}\equiv
\overline q_3\,q_2\ .$$

\begin{figure}[b]\centering\includegraphics[scale=.91687,clip]{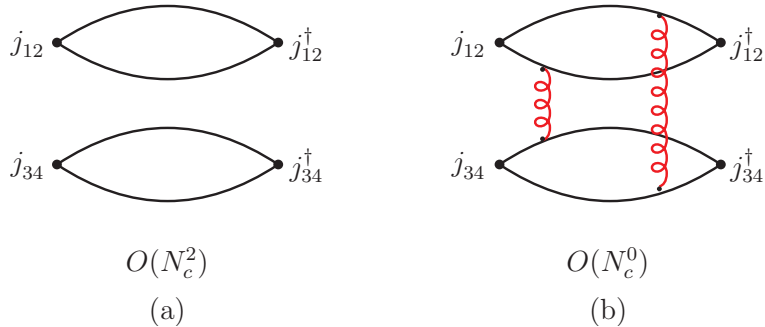}
\caption{Examples of the (a) leading and (b) subleading Feynman
diagrams for \emph{direct\/} Green functions $\Gamma^{\rm(dir)}
_{\rm I}$ \cite[Fig.~1]{TQN2}; similar Feynman diagrams exist for
\emph{direct\/} Green functions $\Gamma^{\rm(dir)}_{\rm II}$.
Curly red lines indicate~gluons.}\label{f1}\end{figure}

\begin{figure}[t]\centering\includegraphics[scale=.91687,clip]{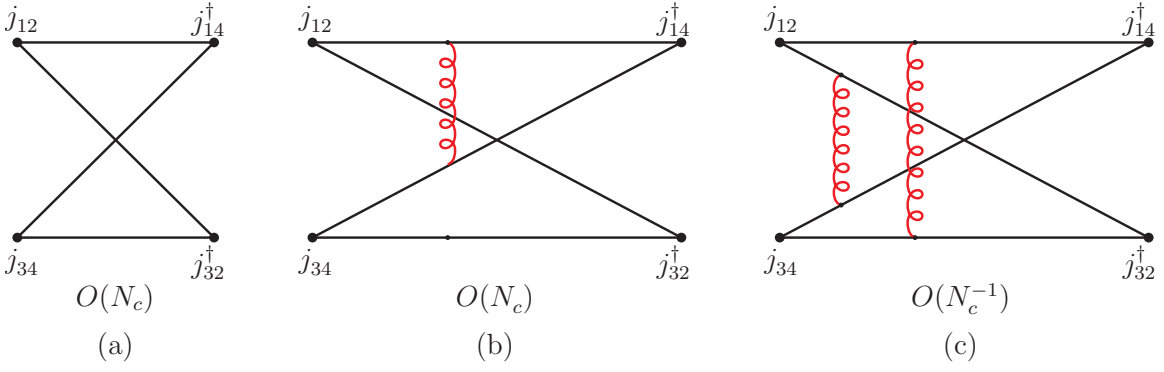}
\caption{Typical representatives of the (a) leading or (b,c)
subleading Feynman diagrams for \emph{recombination\/} Green
functions $\Gamma^{\rm(recomb)}$ \cite[Fig.~2(a,b,e)]{TQN2}; as in
Fig.~\protect\ref{f1}, curly red lines indicate internal exchanges
of gluons.}\label{f2}\end{figure}

In order to ascertain that a given QCD diagram may contain a
tetraquark contribution in form of a pole term, we have to ensure
that it receives a four-quark contribution to its $s$-channel
singularities, in addition to gluon singularities that do not
modify that diagram's $N_{\rm c}$ behaviour. If this tetraquark is
composed of two quarks and two antiquarks with masses $m_j$,
$j=1,2,3,4$, each diagram in question should develop, as a
function of the Mandelstam variable $s$, a four-particle branch
cut, \pagebreak starting~at~the branch point
$s=(m_1+m_2+m_3+m_4)^2$ \cite{TQN1,TQN2}. The latter's existence
can be easily checked by use of the Landau equations \cite{LDL}.
Diagrams that exhibit no $s$-channel singularity at all or only
two-particle (quark--antiquark) singularities cannot contribute to
the formation of tetraquarks at $N_{\rm c}$-leading order and thus
should not be taken into account for the $N_{\rm c}$-behaviour
analysis of any~tetraquark properties. With regard to the
allocation of the four unequal quark flavours in initial- and
final-state mesons, we encounter two types of scattering
reactions, \emph{viz.}, ``direct'' processes, where the flavour
distribution is preserved, and ``recombination'' processes, where
the flavour distribution undergoes rearrangement:
\begin{align*}M_{12}+M_{34}&\longrightarrow M_{12}+M_{34}&&
\mbox{(direct channel I)}\ ,\\M_{14}+M_{32}&\longrightarrow
M_{14}+M_{32}&&\mbox{(direct channel II)}\ ,\\M_{12}+M_{34}&
\longrightarrow M_{14}+M_{32}&&\mbox{(recombination channel)}\
.\end{align*}For both of the ``direct'' reactions, we scrutinize
\cite{TQN1,TQN2,TQN3} the two four-current Green functions
(Fig.~\ref{f1})$$\Gamma^{\rm(dir)}_{\rm I}\equiv\langle T(j_{12}\,
j_{34}\,j^\dag_{12}\,j^\dag_{34})\rangle\ ,\qquad\Gamma^{\rm(dir)}
_{\rm II}\equiv\langle T(j_{14}\,j_{32}\,j^\dag_{14}\,j^\dag_{32})
\rangle\ .$$Only Feynman diagrams of the categories represented by
Fig.~\ref{f1}(b) may receive contributions from a tetraquark; the
large-$N_{\rm c}$ behaviour of the two sets \emph{potentially\/}
supporting tetraquarks (T) is the~same:$$\Gamma^{\rm(dir)}_{\rm
I,T}=O(N_{\rm c}^0)\ ,\qquad\Gamma^{\rm(dir)}_{\rm II,T}=O(N_{\rm
c}^0)\ .$$Likewise, for the ``recombination'' reaction we study
\cite{TQN1,TQN2,TQN3} the four-current Green function
(Fig.~\ref{f2})$$\Gamma^{\rm(recomb)}\equiv\langle T(j_{12}\,
j_{34}\,j^\dag_{14}\,j^\dag_{32})\rangle\ .$$Only Feynman diagrams
of the type of Fig.~\ref{f2}(c) may have tetraquark poles,
behaving at large~$N_{\rm c}$~like$$\Gamma^{\rm(recomb)}_{\rm
T}=O(N_{\rm c}^{-1})\ .$$

\begin{figure}[ht]\centering\includegraphics[scale=.91687,clip]{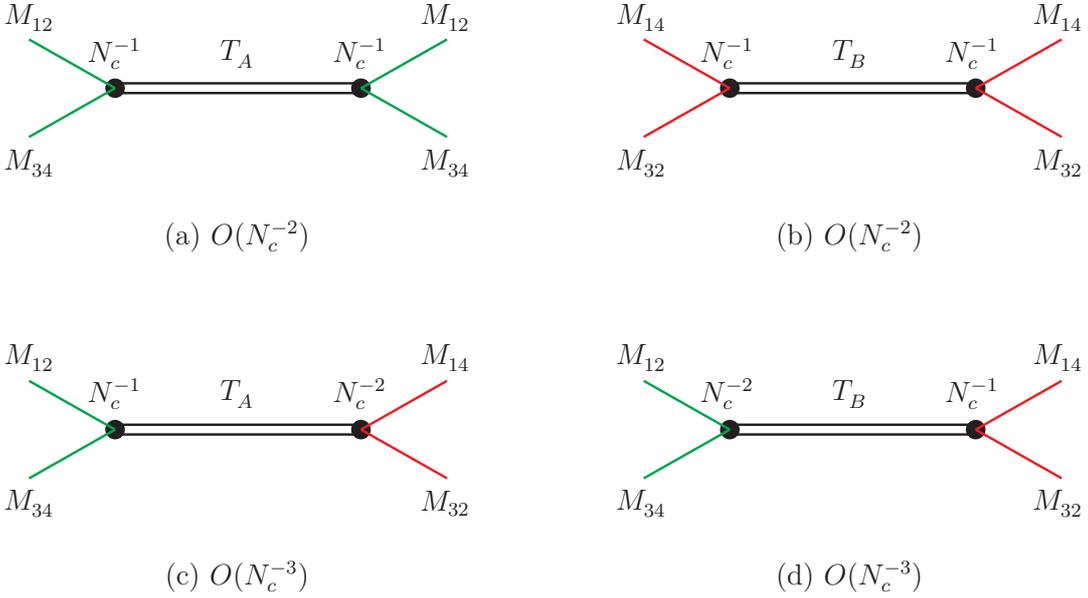}
\caption{Meson--meson scattering amplitudes: $N_{\rm c}$-leading
contributions at two tetraquark poles \cite[Fig.~7]{TQN2}.}
\label{f3}\end{figure}

\noindent Since the large-$N_{\rm c}$ behaviours of the pole
contribution to direct and recombination correlators
differ,$$\Gamma^{\rm(dir)}_{\rm I,T}=O(N_{\rm c}^0)\ ,\qquad
\Gamma^{\rm(dir)}_{\rm II,T}=O(N_{\rm c}^0)\ ,\qquad
\Gamma^{\rm(recomb)}_{\rm T}=O(N_{\rm c}^{-1})\ ,$$reproducing
\cite{TQN1,TQN2} these large-$N_{\rm c}$ findings (Fig.~\ref{f3})
requires (at least) two tetraquarks, say, $T_A$~and~$T_B$, with
different couplings to the ordinary-meson pairs and related
transition amplitudes behaving like\begin{align*}
A(T_A\longleftrightarrow M_{12}\,M_{34})=O(N_{\rm c}^{-1})\ ,
\qquad A(T_A\longleftrightarrow M_{14}\,M_{32})=O(N_{\rm c}^{-2})\
,&\\A(T_B\longleftrightarrow M_{12}\,M_{34})=O(N_{\rm c}^{-2})\ ,
\qquad A(T_B\longleftrightarrow M_{14}\,M_{32})= O(N_{\rm c}^{-1})
\ .&\end{align*}\newpage\noindent From the $N_{\rm c}$-leading
tetraquark--two-ordinary-meson transition amplitudes, the decay
widths of our pair of tetraquarks, $T_A$ and $T_B$, exhibit the
same large-$N_{\rm c}$ behaviour and are definitely narrow
\cite{TQN1,TQN2,TQN3}:$$\Gamma(T_A)=O(N_{\rm c}^{-2})\ ,\qquad
\Gamma(T_B)=O(N_{\rm c}^{-2})\ .$$

The above outcome and the colour structure of the intermediate
states in the Feynman diagrams allow us to offer an educated guess
on the \emph{flavour structure\/} of each of the tetraquarks $T_A$
and $T_B$ \cite{TQN2}. Specifically, colour exchange between the
quarks characterizes the intermediate states. This hints~at$$T_A
\sim(\overline q_1\,q_4)\,(\overline q_3\,q_2)\ ,\qquad T_B\sim
(\overline q_1\,q_2)\,(\overline q_3\,q_4)\ .$$

\section{Implications and Conclusions: Doubtful Existence of
Flavour-Exotic Tetraquarks}
\begin{figure}[ht]\centering\includegraphics[scale=.49,clip]{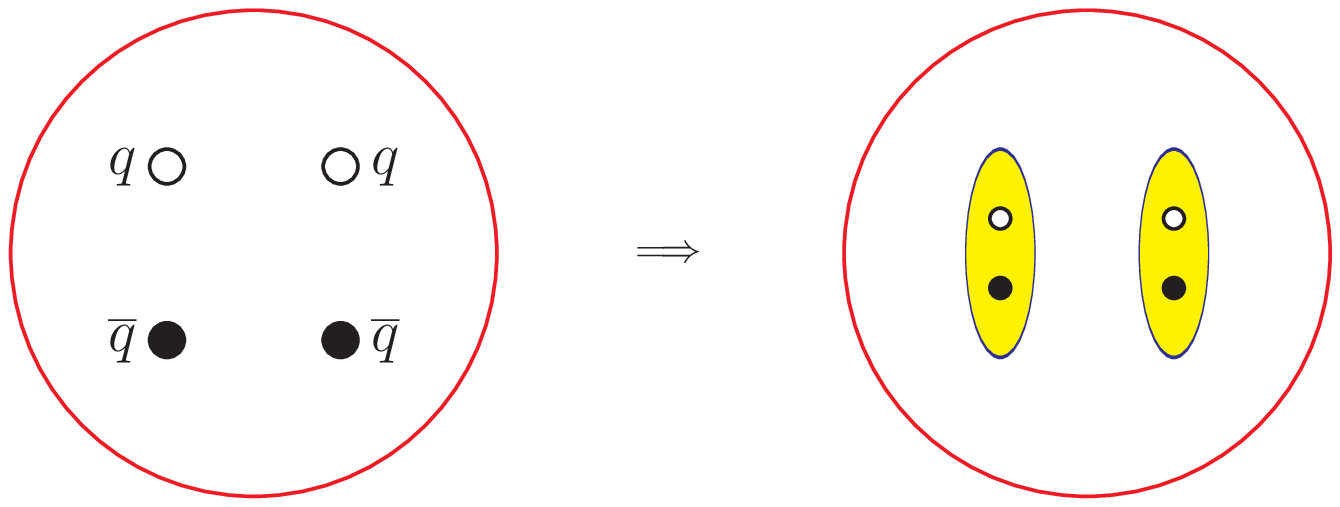}
\hspace{10ex}\includegraphics[scale=.49,clip]{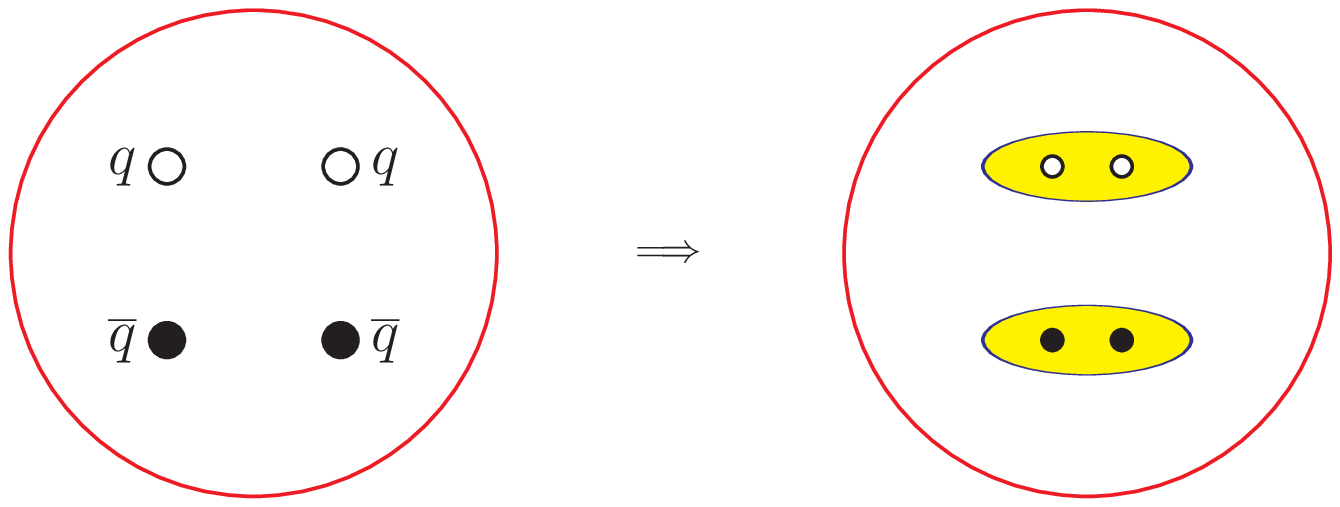}
\caption{Formation of tetraquarks: two-ordinary-meson (left) vs.\
diquark--antidiquark (right) configuration.}\label{f910a}
\end{figure}

These insights tend to favour a
singlet--singlet colour structure (Fig.~\ref{f910a}) of the two
flavour-exotic tetraquark companions, perhaps with mixings of
order $O(1/N_{\rm c})$ of the two configurations \cite{TQN2}.
The\pagebreak

\noindent mutual interactions of the two colour-singlet
clusters are no longer confining: the four-quark system is no
longer compact but resembles a \emph{molecular\/} state. Compact
multiquarks result from the \emph{diquark\/} mechanism
(Fig.~\ref{f910a}) \cite{JW,SZ,MPPR}: Confining interactions bind
colour-nonsinglet diquark and antidiquark clusters to compact
states; attractiveness compels diquarks to form in the sole
colour-antisymmetric representation. That uniqueness is at odds
with the $N_{\rm c}$-driven need for two unequal tetraquarks
\cite{TQN3}. Hidden dynamical mechanisms \cite{BHL,MPR} can avoid
quark--antiquark over diquark forces' dominance. Molecular-like
tetraquarks are tougher: Typical effective meson--meson couplings
are of order $N_{\rm c}^{-1}$. Thus, bound-state or resonance
formation requires to sum diagrams of different $N_{\rm c}$
behaviours \cite{P}.

\vspace{1ex}\noindent{\scriptsize{\bf Acknowledgements.} D.~M.~is
supported by the Austrian Science Fund (FWF), Project No.~P29028,
H.~S.~by EU research and innovation\\[-1ex] programme Horizon
2020, grant agreement No.~824093, and D.~M.~and H.~S.~by joint
CNRS/RFBR Grant No.~PRC \mbox{Russia/19-52-15022.}}

\end{document}